\documentclass[preprint,12pt]{elsarticle}

\usepackage{graphicx}

\usepackage{amssymb}

\journal{Physics Letters B}

\begin{document}

\begin{frontmatter}



\title{Measurement of zero degree single photon energy spectra for $\sqrt{s}$ = 7\,TeV proton-proton collisions at LHC}


\author[firenze-infn,firenze-univ]{O. Adriani}
\author[firenze-infn]{L. Bonechi}
\author[firenze-infn]{M. Bongi}
\author[firenze-infn,firenze-univ]{G. Castellini}
\author[firenze-infn,firenze-univ]{R. D'Alessandro}
\author[spain]{A. Faus}
\author[nagoya]{K. Fukatsu}
\author[polyteque]{M. Haguenauer}
\author[nagoya,kmi]{Y. Itow}
\author[waseda]{K. Kasahara}
\author[nagoya]{K. Kawade}
\author[cern]{D. Macina}
\author[nagoya]{T. Mase}
\author[nagoya]{K. Masuda}
\author[nagoya]{Y. Matsubara}
\author[firenze-infn,kmi]{H. Menjo}
\author[nagoya]{G. Mitsuka}
\author[nagoya]{Y. Muraki}
\author[waseda]{M. Nakai}
\author[catania-infn]{K. Noda}
\author[firenze-infn]{P. Papini}
\author[cern]{A.-L. Perrot}
\author[firenze-infn,csfn]{S. Ricciarini}
\author[nagoya,kmi]{T. Sako\corref{cor1}}
\author[waseda]{Y. Shimizu}
\author[nagoya]{K. Suzuki}
\author[waseda]{T. Suzuki}
\author[nagoya]{K. Taki}
\author[kanagawa]{T. Tamura}
\author[waseda]{S. Torii}
\author[catania-infn,catania-univ]{A. Tricomi}
\author[lbn]{W. C. Turner}
\author[spain]{J. Velasco}
\author[firenze-infn]{A. Viciani}
\author[shibaura]{K. Yoshida}

\address[firenze-infn]{INFN Section of Florence, Italy}
\address[firenze-univ]{University of Florence, Italy}
\address[csfn]{Centro Siciliano di Fisica Nucleare e Struttura della Materia, Catania, Italy}
\address[nagoya]{Solar-Terrestrial Environment Laboratory, Nagoya University, Nagoya, Japan}
\address[kmi]{Kobayashi-Maskawa Institute for the Origin of Particles and the Universe, Nagoya University, Nagoya, Japan}
\address[polyteque]{Ecole-Polytechnique, Palaiseau, France}
\address[waseda]{RISE, Waseda University, Japan}
\address[cern]{CERN, Switzerland}
\address[kanagawa]{Kanagawa University, Japan}
\address[catania-infn]{INFN Section of Catania, Italy}
\address[catania-univ]{University of Catania, Italy}
\address[lbn]{LBNL, Berkeley, California, USA}
\address[shibaura]{Shibaura Institute of Technology, Japan}
\address[spain]{IFIC, Centro Mixto CSIC-UVEG, Spain}

\cortext[cor1]{sako@stelab.nagoya-u.ac.jp}

\begin{abstract}
In early 2010, the Large Hadron Collider forward (LHCf) experiment measured very forward 
neutral particle spectra in LHC proton-proton collisions.
From a limited data set taken under the best beam conditions (low beam-gas background and
low occurance of pile-up events), the single photon spectra at $\sqrt{s}$=7$\,$TeV and 
pseudo-rapidity
($\eta$) ranges from 8.81 to 8.99 and from 10.94 to infinity were obtained for the first 
time and are reported in this paper.
The spectra from two independent LHCf detectors are consistent with one another and serve 
as a cross check of the data.
The photon spectra are also compared with the predictions of several hadron interaction
models that are used extensively for modeling ultra high energy cosmic ray showers.
Despite conservative estimates for the systematic errors, none of the models
agree perfectly with the measurements.
A notable difference is found between the data and the DPMJET 3.04 and PYTHIA 8.145 
hadron interaction models above 2$\,$TeV where the models predict higher photon yield than
the data.
The QGSJET II-03 model predicts overall lower photon yield than the data, especially
above 2$\,$TeV in the rapidity range 8.81$<$$\eta$$<$8.99.
\end{abstract}


\end{frontmatter}


\section{Introduction} \label{sec-intro}
The lack of knowledge about forward particle production in hadron collisions 
affects the interpretation of observations of Ultra-High Energy Cosmic-Rays 
(UHECR).
Although UHECR observations have made notable improvements in the last few years
\cite{ref-AugerSpectrum} \cite{ref-AugerAniso} \cite{ref-AugerComp} 
\cite{ref-HiResSpectrum} \cite{ref-HiResAniso} \cite{ref-HiResComp} \cite{ref-TA},
some critical parts of the analysis depend on the Monte Carlo (MC) simulations of
air shower development that are sensitive to the choice of the hadron interaction 
model.
Accelerator data on the production of very forward emitted particles are indispensable
for constraining the hadron interaction models but are usually not available from the
large general purpose detectors.
The Large Hadron Collider forward (LHCf) experiment has been designed to measure the
neutral particle production cross sections at very forward collision angles of LHC
proton-proton collisions, including zero degrees.
When the LHC reaches its designed goal of 14$\,$TeV collision energy, the energy 
in the equivalent laboratory frame will be 10$^{17}$$\,$eV, a factor of one thousand
increase compared to previous accelerator data in the very forward regions 
\cite{ref-ISR} \cite{ref-UA7}.

Two detectors, called Arm1 and Arm2, have been installed in the instrumentation 
slots of the TANs (Target Neutral Absorbers) located $\pm$140$\,$m from the ATLAS 
interaction point (IP1) and at zero degree collision angle.
Inside a TAN the beam vacuum chamber makes a Y shaped transition from a single 
common beam tube facing the IP to two separate beam tubes joining to the arcs 
of LHC. 
Charged particles from the IP are swept aside by the inner beam separation 
dipole D1 before reaching the TAN so only neutral particles are incident on the LHCf
detectors. 
This unique location covers the pseudo-rapidity range from 8.7 (8.4 in case of the
operation with the maximum beam crossing angle) to infinity (zero degrees).
Each detector has two sampling and imaging calorimeters composed of 44 radiation 
lengths (1.55 hadron interaction lengths) of 
tungsten and 16 sampling layers of 3$\,$mm thick plastic scintillators.
The transverse sizes of the calorimeters are 20$\,$mm$\times$20$\,$mm and 
40$\,$mm$\times$40$\,$mm in Arm1, and 25$\,$mm$\times$25$\,$mm and 
32$\,$mm$\times$32$\,$mm in Arm2.
The smaller calorimeters cover the zero degree collision angle.
Four X-Y layers of position sensitive detectors (scintillating fiber, SciFi, 
belts in Arm1 and silicon micro-strip sensors in Arm2; 1$\,$mm and 0.16$\,$mm 
readout pitchs, respectively) are inserted in order to 
provide transverse positions of the showers.
The LHCf detectors have energy and position resolutions for the 
electromagnetic showers better than 5\% and 200\,$\mu$m, respectively, in the
energy range $>$100$\,$GeV.
More detail on the scientific goals, construction and performance of the
detectors can be found in previous reports
\cite{ref-LHCfTDR} \cite{ref-LHCfJINST} \cite{ref-prototype} \cite{ref-sps2007} 
\cite{ref-LHCfsilicon} \cite{ref-menjopi0}.

This paper describes the first analysis results of LHCf data.
Single photon energy spectra are reported for $\sqrt{s}$ = 7$\,$TeV proton-proton
collisions.
In Sec.\ref{sec-data} the data set used in the analysis is introduced.
In Sec.\ref{sec-anal} the analysis process and experimental results are presented.
Beam related background and uncertainties are discussed
in Sec.\ref{sec-bg}.
The experimental results are compared with MC predictions of several hadron
interaction models in Sec.\ref{sec-comp} and summarized in Sec.\ref{sec-disc}.

\section{Data} \label{sec-data}
\begin{figure}
  \begin{center}
    \includegraphics[width=140mm]{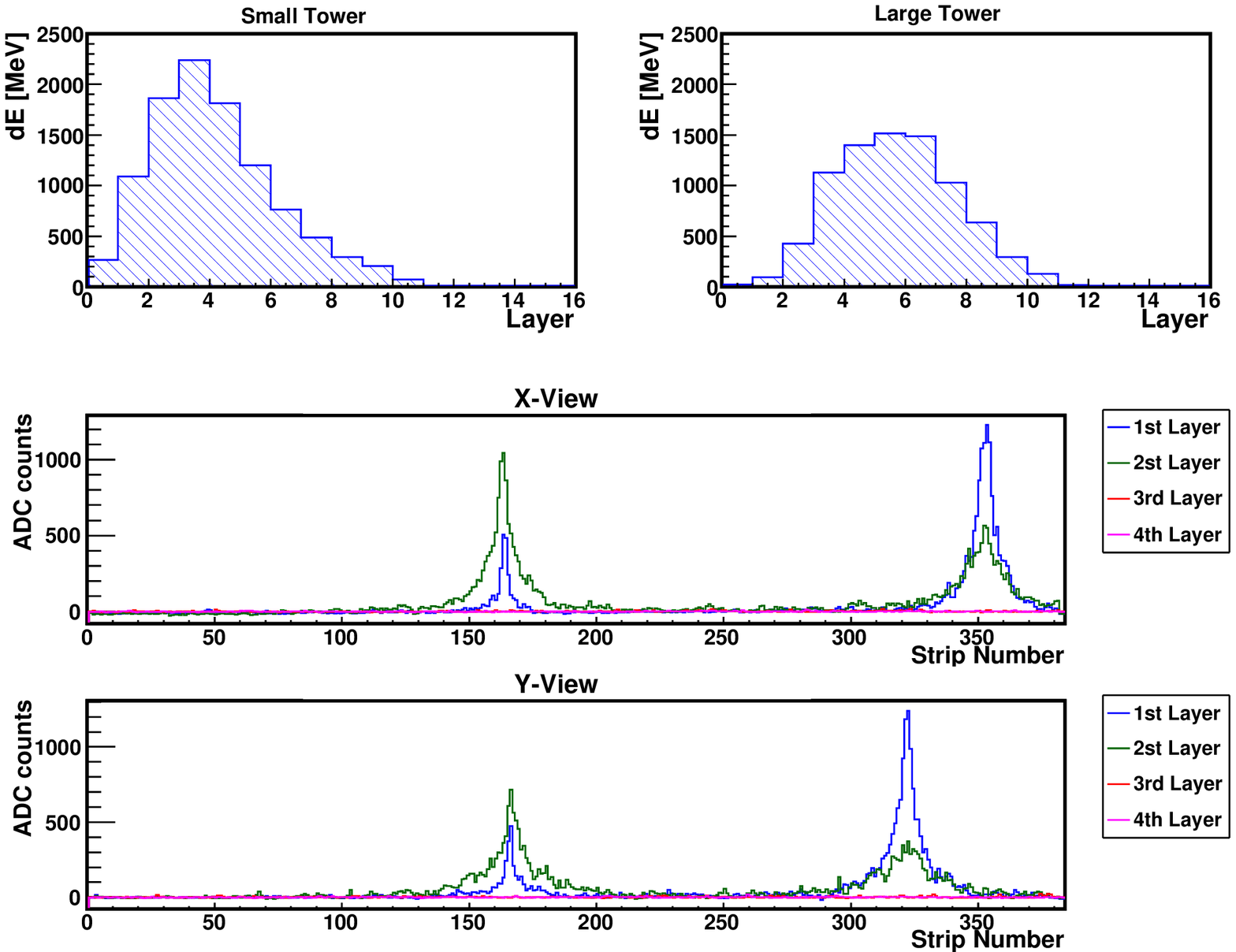}
    \caption{An example of the $\pi^{0}$ candidate events observed in the LHCf Arm2 detector.
             Top two figures show the longitudinal developments of the two photon initiated
             showers observed in the 25$\,$mm and the 32$\,$mm calorimeters.
             In the middle and bottom panels, transverse X and Y profiles of the showers
             are shown.
             Different colors indicate data in the different silicon layers.
             }
    \label{fig-event}
  \end{center}
\end{figure}

Data used in this analysis was obtained on 15 May 2010 during proton-proton
collisions at $\sqrt{s}$=7$\,$TeV with zero degree beam crossing angle (LHC Fill 1104).
The total luminosity of the three crossing bunches in this fill, 
L=(6.3--6.5)$\times$10$^{28}$$\,$$cm^{-2}s^{-1}$, 
provided ideal operating conditions as discussed in Sec.\ref{sec-bg}.
The data that were taken during a luminosity optimization scan were eliminated from 
the analysis.
The trigger for LHCf events was generated at three levels.
The first level trigger (L1T) was generated from beam pickup signals (BPTX) when
a bunch passed IP1.
A shower trigger was generated when signals from any successive 3 scintillation 
layers in any calorimeter exceeded a predefined threshold.
Then the second level trigger for shower events (L2TA) was issued when the data 
acquisition system was armed.
The threshold was chosen to achieve $>$99\% efficiency for $>$100$\,$GeV photons. 
Data were recorded with the third level trigger (L3T) when all the other types of 
second level triggers (pedestal, laser calibration, etc) were combined.
Examples of the longitudinal and lateral development of electromagnetic
showers observed in the Arm2 detector are shown in Fig.\ref{fig-event}.
In this case two electromagnetic showers from $\pi^{0}$ decay into two photons are shown,
with each photon striking a different calorimeter of the Arm2 detector.
The generation of the L2TA and L3T triggers, and hence the data recording, were performed
independently for the Arm1 and Arm2 detectors.
Data acquisition was carried out under 85.7\% (Arm1) and 67.0\% (Arm2) average livetimes 
($\epsilon_{DAQ}$).
The livetimes were defined as $\epsilon_{DAQ}$ = N$_{L2TA}$/N$_{shower}$ where N$_{shower}$ 
and N$_{L2TA}$ are the number of counts in the shower and L2TA triggers, respectively.

The integrated luminosities ($\int$$\,$L$\,$dt) corresponding to the data used in this 
paper are 0.68$\,$nb$^{-1}$ (Arm1) and 0.53$\,$nb$^{-1}$ (Arm2) after the data taking
livetimes are taken into account.
The absolute luminosity is derived from the counting rate of the Front Counters (FC)
\cite{ref-LHCfJINST}.
FCs are thin plastic scintillators fixed in front of the LHCf main calorimeters and
covering a wide aperture of 80$\,$mm$\times$80$\,$mm.
The calibration of the FC counting rates to the absolute luminosity was made during
the Van der Meer scans on 26 April and 9 May 2010.
The calibration factors obtained from the two scans differ by 2.1\%.
The estimated luminosities for the 15 May data differ by 2.7\% between the two FCs.
Considering the uncertainty of $\pm$5.0\% in the beam intensity measurement during 
the Van der Meer scans
\cite{ref-bcnwg-note}, we estimate an uncertainty of $\pm$6.1\% in the luminosity 
determination.

\section{Analysis and Results} \label{sec-anal}
\subsection{Event Reconstruction} \label{sec-recon}
The same analysis process has been adapted to each Arm independently.
The transverse impact position of a particle is determined using 
the information provided by the position sensitive detectors.
Using the position information, the raw data from the scintillation layers are 
corrected for the non-uniformity of light collection and for shower `leakage out' the 
edges of the layers \cite{ref-prototype}.
Events that fall within 2$\,$mm of the edges are removed from the analysis due to the
large uncertainty in the energy determination owing to shower leakage.
The recorded charge information is converted to deposited energy
based on calibration runs with SPS fixed target experiments below 200$\,$GeV 
\cite{ref-LHCfTDR} \cite{ref-sps2007}.
The sum of the energy deposited in the 2$^{nd}$--13$^{th}$ scintillation layers is
converted to the primary photon energy by using a function determined by MC 
simulation using the EPICS 8.81/COSMOS 7.49 simulation package \cite{ref-EPICS}
and confirmed in the SPS beam tests.
Note that this energy estimate does not represent the incident energy of hadrons
because our calorimeters have only 1.55 hadron interaction lengths.
In the detector simulations the Landau-Pomeranchuk-Migdal effect 
\cite{ref-LPMa} \cite{ref-LPMb} has been
considered and neglecting the LPM effect does not change the energy estimate
at the 1\% level because we sum the deposited energy up to a sufficiently deep layer.

The linearity of each PMT was carefully tested before detector assembly over a wide 
range of signal amplitude by exciting a scintillator using a 337$\,$nm UV laser pulse 
\cite{ref-LHCfTDR} \cite{ref-prototype}.
Although the measured non-linear response functions have been applied in the analysis,
the difference between linear and non-linear reconstructions for 3$\,$TeV photons is
only 0.5\% at maximum.
We also took data under LHC conditions with different PMT gains, but after applying
gain calibrations no difference in the data sets was observed.
Events having energy below 100$\,$GeV are eliminated from the analysis to avoid
corrections due to the trigger inefficiency and to reject particles produced in the
interaction between collision products and the beam pipe.

\subsection{Single Event Selection}
To deduce the single photon energy, multi-hit events with more than one photons
registered in a single calorimeter are eliminated.
These multi-hit events are identified by using the lateral shower distribution
measured by the position sensitive layers.
According to MC simulation, the efficiency for correctly identifying true single
photon events is $>$98\%.
The efficiency for identifying multi-hit events depend on the distance and the energy
ratio of two photons and the detectors because of the different readout pitchs of
the Arm1 SciFi belts and the Arm2 silicon micro-strip sensors.
When the separation is greater than 1$\,$mm and the lower energy photon has more than
5\% of the energy of the nearby photon, the efficiencies for identifying multi-hit
events are $>$70\% and $>$90\% for Arm1 and Arm2, respectively. 

To estimate the systematic uncertainty in the multi-hit identification efficiency,
we produced an artificial sample of multi-hit event sets by superimposing two clearly
single photon-like events for both the experimental and MC data based on the EPOS 1.99
model.
Details of the MC simulations are described in Sec.\ref{sec-comp}.
To choose the energies and separation of a photon pair, we followed the distributions
determined by the DPMJET 3.04 model.
For the two artificial data sets the efficiencies for identifying multi-hit events
do not differ by more than 10\% and 3\% over the entire energy range for Arm1 and Arm2,
respectively.
This affects the final single photon energy spectrum shape by less than 1\% below 
1.5$\,$TeV and increasingly up to 2--20\% at 3$\,$TeV.
Tha maximum difference is found for the Arm1 large calorimeter.

Next, we compared the effect of the multi-hit cut on the Arm1 and Arm2 detectors.
While the fraction of events thrown out by this cut differs by less than 5\% between 
0.5$\,$TeV and 1.5$\,$TeV, it gradually increases to 30\% and 60\% at 3$\,$TeV
for the small and large calorimeters, respectively.
The main reason for these differences is the different geometry of the Arm1 and Arm2 
calorimeters.
The different performances of the position sensitive detectors in the two Arms
and an uncertainty in the absolute energy scale discussed in Sec.\ref{sec-pi0}
may also contribute to the differences in multi-hit identification fractions of the
Arm1 and Arm2 detectors.
Because we cannot presently separate the sources of the difference and hence cannot
apply corrections to the data, we assign the differences divided by $\sqrt{2}$ as part
of the systematic uncertainty for each detector.
Finally we take quadratic sum of the two uncertainties related to the multi-hit 
identification efficiency and the multi-hit cut as systematic error of the
single photon selection procedure.

\subsection{Photon Event Selection}
\begin{figure}
  \begin{center}
    \includegraphics[width=120mm]{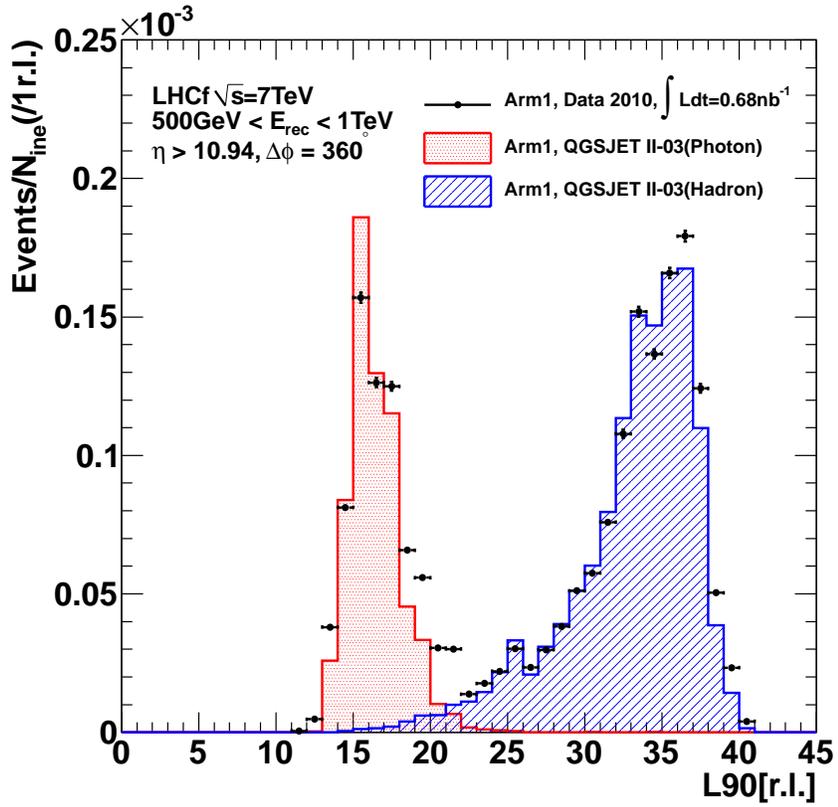}
    \caption{
             The L$_{90\%}$ distribution measured by the Arm1 20$\,$mm calorimeter
             for the reconstructed energy of 500$\,$GeV$~$-$~$1$\,$TeV.
             Plots are experimental data and the red and blue histograms are the
             templates calculated from the pure photon and pure hadron MC events,
             respectively.
             The two templates are independently normalized to best describe the 
             observed data.
             }
    \label{fig-L90}
  \end{center}
\end{figure}

To select only electromagnetic showers and eliminate hadron (predominantly neutron)
contamination, a simple parameter, L$_{90\%}$ is defined.
L$_{90\%}$ is the longitudinal distance in radiation lengths measured from the
entrance to a calorimeter to the position where 90\% of the total shower energy has
been deposited.
Fig.\ref{fig-L90} shows the distribution of L$_{90\%}$ for the 20$\,$mm calorimeter
of the Arm1 detector for the events with the reconstructed energy between 500$\,$GeV
and 1$\,$TeV.
Two distinct peaks are observed corresponding to photon and hadron (neutron) events.
The L$_{90\%}$ distributions for pure photon and hadron samples are generated by MC 
simulation using the collision product generator QGSJET II-03 as shown in 
Fig.\ref{fig-L90}.
They are called `templates' hereafter.
The choice of hadron interaction model in determining the template does not
affect the results in this paper. 
In the event selection, we set an energy dependent criteria in L$_{90\%}$ to keep the
photon  detection efficiency $\epsilon_{PID}$=90\% over the entire energy range based 
on the photon template.
The purity of the selected photon events is determined
by normalizing the template functions to the observed L$_{90\%}$ distribution.
The purity, P, is defined as P=N$_{phot}$/(N$_{phot}$+N$_{had}$) in each energy bin.
Here N$_{phot}$ and N$_{had}$ are the numbers of photon and hadron (neutron) events 
in the templates in the selected L$_{90\%}$ range.
Multiplying each energy bin by P$\times$$\epsilon_{PID}^{-1}$, we obtained non biased
photon energy spectra.

Some disagreements in the L$_{90\%}$ distribution are found between the data and the 
MC calculations.
This may be caused by errors in the absolute energy determination and 
channel-to-channel calibrations and may also be motivation for studying the LPM effect
in detail.
Here we consider a systematic uncertainty caused by the uncertainty of the 
template fitting method in the correction of the photon spectra.
Small modifications of the template functions, widening with respect to the peak
position up to 20\% and constant shift up to 0.7 radiation lengths, to give the best
match with the data, provide another estimate of the correction to the photon spectra.
The difference of the correction factors between the original and the modified template
methods amount to 5--20\% from low to high photon energy and this is assigned as a
systematic uncertainty of the particle identification in the final spectra.

\subsection{Energy Scale Uncertainty from $\pi^{0}$ Mass Reconstruction} \label{sec-pi0}
When each of two calorimeters records a single photon as shown in Fig.\ref{fig-event}, 
shower `leakage in' is corrected according to a function based on MC simulation.
Using the corrected energies and positions of the shower axes, the invariant mass of 
the photon pair is calculated assuming their vertex is at the interaction point.
In MC simulations of the full detector response and the analysis process, 
we confirmed the reconstructed mass peaks at 135.2$\,$MeV in Arm1 and 135.0$\,$MeV
in Arm2, thus reproducing the $\pi^{0}$ mass.
The statistical uncertainty in the reconstructed invariant mass of the MC simulations
is $\pm$0.2$\,$MeV.

On the other hand, the reconstructed invariant masses of photon pairs for the experimental
data are 
145.8$\pm$0.1$\,$MeV (Arm1) and 140.0$\pm$0.1$\,$MeV (Arm2) where $\pm$0.1$\,$MeV
uncertainties are statistical.
A portion of the 7.8\% and 3.7\% invariant mass excess compared to the $\pi^{0}$ mass
reconstructed in the MC simulations can be explained by the well understood systematic
error of the absolute energy scale, estimated to be $\pm$3.5\%.
This 3.5\% systematic error is dominated by the errors in factors converting measured
charge to deposited energy and by the errors in corrections for non-uniform light
collection efficiency.
Uncertainties in determining the opening angle of a photon pair and the shower 
leakage-in correction, typically $\pm$1\% and $\pm$2\% respectively, are also sources
of error in mass reconstruction.
These known elements quadratically add up to a systematic mass shift of
4.2\% and can explain the mass shift in the Arm2 detector, but not Arm1.

Because all the two photon invariant mass shift may not be due to the energy scale 
uncertainty,  we did not apply any correction for energy scale in the energy spectra
presented in Sec.\ref{sec-spectra}.
Instead we assigned asymmetric systematic errors in the absolute energy scale,
[-9.8\%, +1.8\%] (Arm1) and [-6.6\%, +2.2\%] (Arm2).
Here we assumed uniform and Gaussian probability distributions for the energy scale
errors estimated from the mass shift (7.8\% and 3.7\%) and the known systematics (3.5\%),
respectively.
After the standard deviations of two components (7.8\%/$\sqrt{3}$ in case of the
uniform probability distribution) were quadratically added, the systematic error
bands are assigned with respect to the central value of the mass shift. 
To determine the systematic errors in the final energy spectra, we reconstructed two
energy spectra by scaling the energy using the two extremes quoted above.
The differences from the non-scaled spectrum to the two extreme spectra are assigned
as systematic errors in each energy bin.

\subsection{Spectra Reconstruction} \label{sec-spectra}
To compensate for the different geometry of two arms, we selected common rapidity
and azimuthal ranges to deduce the photon energy spectra.
The ranges used for the small calorimeters and the large calorimeters are 
[$\eta$$>$10.94, $\Delta\phi$=360.0$^{\circ}$]
and [8.99$>$$\eta$$>$8.81, $\Delta\phi$=20.0$^{\circ}$], respectively.
Here $\eta$, $\phi$ and $\Delta\phi$ represent the pseudo-rapidity, azimuthal direction
and interval of $\phi$ with respect to the beam axis which is centered on the small 
calorimeters.

\begin{figure}
  \begin{center}
    \includegraphics[width=170mm]{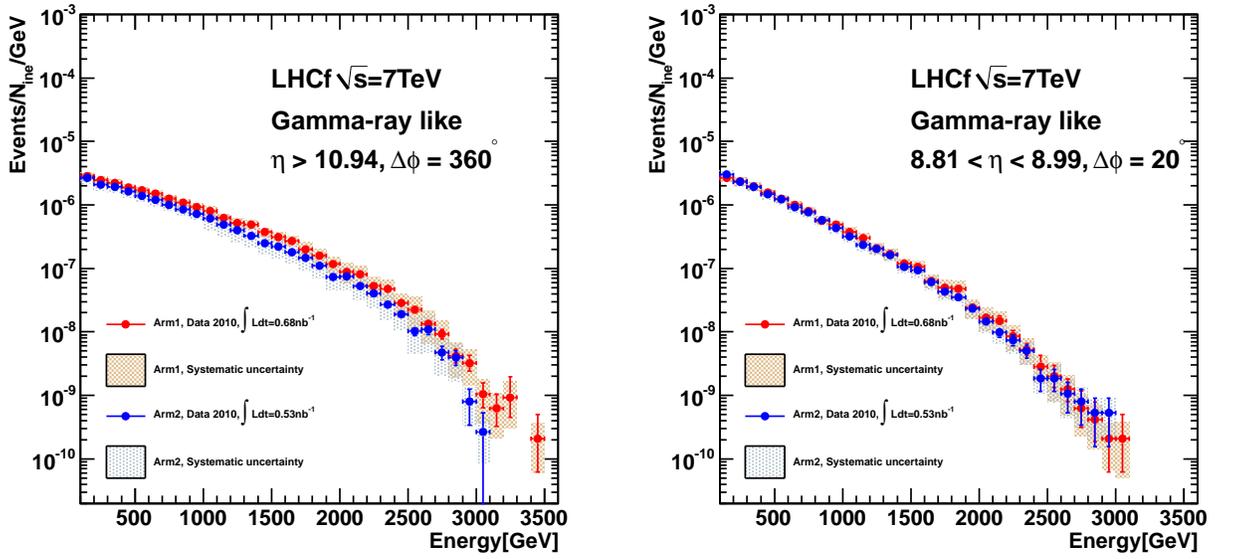}
    \caption{
             Single photon spectra measured by the Arm1 (red) and Arm2 (blue) detectors.
             Left (right) panel shows the results for the small (large) calorimeter
             or large (small) rapidity range.
             The error bars and shaded areas indicate the statistical and systematic
             errors, respectively.
             To discuss consistency of two detectors, only uncorrelated components 
             are plotted for the systematic errors.
             }
    \label{fig-gamma}
  \end{center}
\end{figure}

Photon spectra measured in the small and large calorimeters are
shown in Fig.\ref{fig-gamma}.
The red and blue plots show the results from Arm1 and Arm2, respectively.
The error bars and shaded areas indicate the one standard deviation statistical and
systematic errors, respectively, uncorrelated between the two detectors.
On the vertical axis, the number of inelastic collisions, N$_{ine}$, is calculated
as N$_{ine}$ = $\sigma_{ine}$ $\int$$\,$L$\,$dt assuming the inelastic cross 
section $\sigma_{ine}$ = 71.5$\,$mb.
Using the integrated luminosities introduced in Sec.\ref{sec-data},
N$_{ine}$ = 4.9$\times$10$^{7}$ for Arm1 and 3.8$\times$10$^{7}$ for Arm2.
From Fig.\ref{fig-gamma} we find general agreement between the two Arms that are within
the errors.
The reason for the difference between the two Arms in the small calorimeters 
(higher rapidity) is not yet understood.
However, because the difference is still within the errors, we did not apply any
correction.

Here we note that the obtained spectra are expected to be distorted from the
single photon inclusive spectra due to the analysis processes especially the 
multi-hit cut that reflects the physical size of the LHCf detector and the differences
between the hadron interaction models.
The multi-hit cut is expected to suppress the event rate per pp interaction while
the inefficiency of multi-hit identification raises the event rate at high energy
because we misreconstruct multi photon energy as single photon energy.
To estimate the deformation of the energy spectra the reconstructed energy spectra of
photons normalized to
the true inclusive single photon spectra are studied by MC simulations for the
different hadron interaction models.
Here in calculating the `true inclusive single photon spectra' particle decay in the 
140$\,$m flight path from the interaction point to the LHCf detectors and the LHCf
calorimeter aperture are taken into account.
In the case of multi-hits in a single calorimeter, each photon is counted independently.  
As a result, 0--15$\%$ of energy independent supression is found below 2$\,$TeV and it
turns to gradually rise up to +15$\%$ over 3$\,$TeV.
The maximum difference between interaction models and between the two
arms are about 10$\%$ and 5$\%$, respectively.

\section{Beam Related Background and Uncertainties} \label{sec-bg}
The events containing more than one collision (pile-up events) in a single bunch
crossing may cause an additional bias. 
Given that a collision has occurred, the probability of pile-up (P(n$\ge$2)/P(n$\ge$1))
can be calculated from the Poisson probability distribution.
Using the highest bunch luminosity of L=2.3$\times$10$^{28}$ cm$^{-2}$s$^{-1}$ 
used in this analysis, inelastic cross section $\sigma_{ine}$ = 71.5$\,$mb
and the revolution frequency of LHC f$_{rev}$ = 11.2$\,$kHz, the probability is
P(n$\ge$2)/P(n$\ge$1)=0.072.
Considering the acceptance of a LHCf calorimeter for an inelastic collision, 
$\sim$0.03, only 0.2\% of events have more than one event due to the pile-up and 
they are eliminated in the multi-hit cut.
We conclude that pile-up does not affect our analysis.

In the geometrical analysis of the data, we assumed the projected position of the 
zero degree collision angle at the LHCf detectors, referred to as the `beam-center' 
hereafter, can move from fill to fill owing to slightly different beam transverse position
and crossing angles at the IP.
We determined the `beam center' at the LHCf detectors by two methods; first by using the
distribution of particle impact positions measured by the LHCf detectors and second by 
using the information from the Beam Position Monitors (BPMSW) installed $\pm$21$\,$m from
the IP.
From the analysis of the fills 1089--1134,
we found a maximum $\sim$4$\,$mm shift of the beam center at the LHCf detectors,
corresponding to a crossing angle of $\sim$30$\,$$\mu$rad assuming the beam transverse
position did not change.
The two analyses gave consistent results for the location of the beam center on the 
detectors within 1$\,$mm accuracy.
In the geometrical construction of events we used the beam-center determined by LHCf data.
We derived photon energy spectra by shifting the beam-center by 1$\,$mm.
The spectra are modified by 5-20\% depending on the energy and the rapidity range.
This is assigned as a part of systematic uncertainty in the final energy spectra.

The background from collisions between the beam and the residual gas in the vacuum
beam pipe can be estimated from the data.
During LHC operation, there were always bunches that did not have a colliding bunch
in the opposite beam at IP1.
We call these bunches `non-crossing bunches' while the normal bunches are called as
`crossing bunches.'
The events associated with the non-crossing bunches are purely from the beam-gas background
while the events with the crossing bunches are mixture of beam-beam collisions and beam-gas
background. 
Because the event rate of the beam-gas background is proportional to the bunch intensity,
we can calculate the background spectrum contained in the crossing bunch data by scaling 
the non-crossing bunch events.
We found the contamination from the beam-gas background in the final energy spectrum is 
only $\sim$0.1\%.
In addition the shape of the energy spectrum of beam-gas events is similar to that
of beam-beam events, so beam-gas events do not have any significant impact on the
beam-beam event spectrum

The collision products and beam halo particles can hit the beam pipe and produce particles
that enter the LHCf detectors.
However according to MC simulations, these particles have energy below 100$\,$GeV
\cite{ref-LHCfTDR} and do not affect the analysis presented in this paper.

\section{Comparison with Models} \label{sec-comp}
\begin{figure}
  \begin{center}
    \includegraphics[width=180mm]{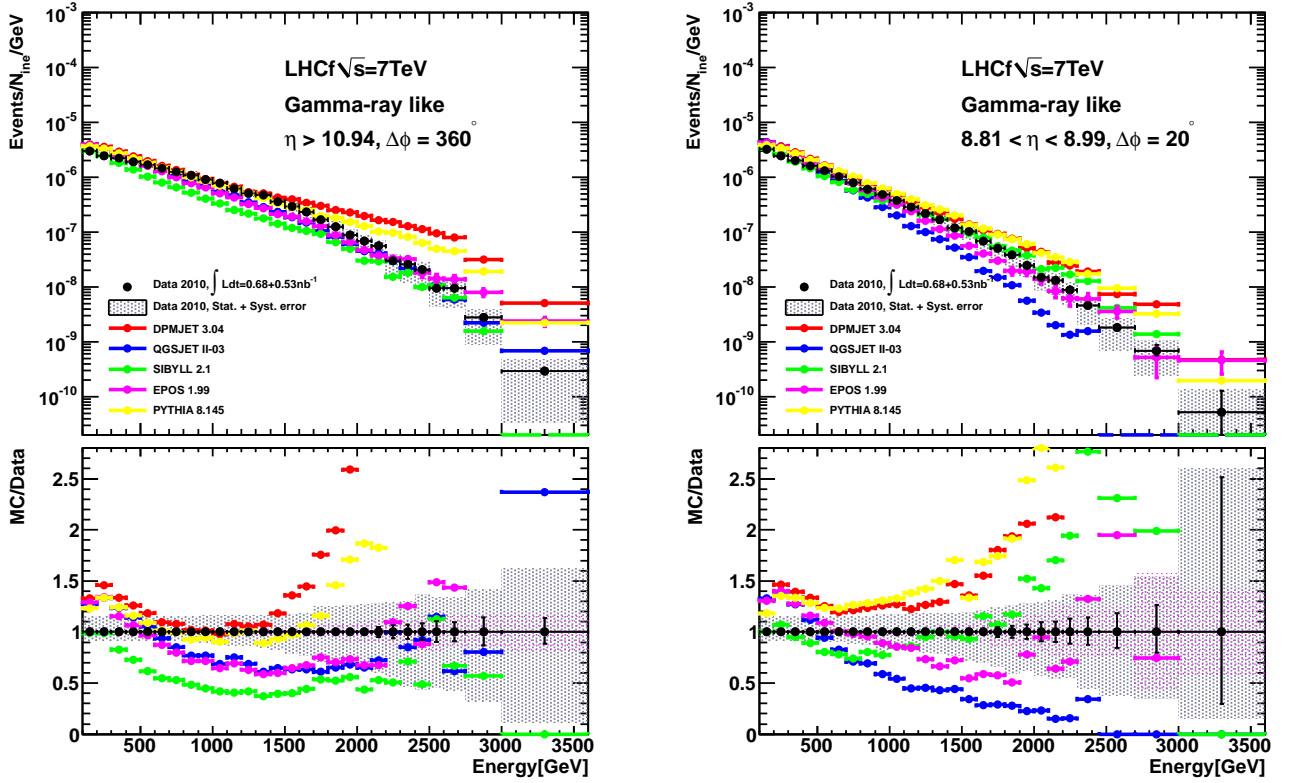}
    \caption{
             Comparison of the single photon energy spectra between the experimental
             data and the MC predictions.
             Top panels show the spectra and the bottom panels show the ratios of
             MC results to experimental data.
             Left (right) panel shows the results for the large (small) rapidity range.
             Different colors show the results from experimental data (black), 
             QGSJET II-03 (blue), DPMJET 3.04 (red), SIBYLL 2.1 (green), 
             EPOS 1.99 (magenta) and PYTHIA 8.145 (yellow).
             Error bars and gray shaded areas in each plot indicate the experimental
             statistical and the systematic errors, respectively.
             The magenta shaded area indicates the statistical error of the MC data set
             using EPOS 1.99 as a representative of the other models.
            }
    \label{fig-compMC}
  \end{center}
\end{figure}

In the top panels of Fig.\ref{fig-compMC} photon spectra predicted by MC 
simulations using different models, 
QGSJET II-03 (blue) \cite{ref-QGS2},
DPMJET 3.04 (red) \cite{ref-DPM3},
SIBYLL 2.1 (green) \cite{ref-SIBYLL},
EPOS 1.99 (magenta) \cite{ref-EPOS}
and PYTHIA 8.145 (default parameter set; yellow) \cite{ref-PYTHIA8a} \cite{ref-PYTHIA8b}
for collisions products are presented together with the combined experimental results.
To combine the experimental data of the Arm1 and Arm2 detectors, the content in each
energy bin was averaged with weights by the inverse of errors.
The systematic uncertainties due to the multi-hit cut, particle identification (PID),
absolute energy scale and beam center uncertainty are quadratically added in each energy
bin and shown as gray shaded areas in Fig.\ref{fig-compMC}.
The uncertainty in the luminosity determination ($\pm$6.1\% as discussed in 
Sec.\ref{sec-data}), that is not shown in Fig.\ref{fig-compMC}, can make an energy 
independent shift of all spectra.

In the MC simulations, 1.0$\times$10$^{7}$ inelastic collisions were generated and 
the secondary particles transported in the beam pipe.
Deflection of charged particles by the D1 beam separation dipole, particle decay and
particle interaction with the beam pipe are taken into account.
The responses of the detectors were calculated using the EPICS/COSMOS libraries
taking the random fluctuation equivalent to electrical noise into account.
The same analysis procedures were then applied to the MC simulations as to the 
experimental data except for the particle identification and its correction.
In the analysis of the MC data set, we used the known information of the particle type.
In the bottom panels the ratios of MC simulations to the experimental data are plotted
together with the statistical and systematic uncertainties of the experimental data.
The statistical uncertainty of the EPOS 1.99 model is also plotted as magenta shaded
areas as a representative of the various models.

We find that none of the models lies within the errors of our data over the entire
energy range.
Some remarkable features are:

1) DPMJET 3.04 and PYTHIA 8.145 show very good agreement with the experimental result
between 0.5 and 1.5$\,$TeV for $\eta$$>$10.94, but they predict significantly larger
photon yield at high energy $>$2$\,$TeV in both rapidity ranges.

2) QGSJET II-03 predicts overall lower photon yield than the experimental result.
This is significant above 2$\,$TeV in the rapidity range 8.81$<$$\eta$$<$8.99.

3) For $\eta$$>$10.94, SIBYLL 2.1 shows a very good agreement with the experimental
result for the spectral shape for $>$0.5$\,$TeV, but predicts a photon yield only
half of the experimental result over the entire energy range.

\section{Summary} \label{sec-disc}
LHCf has measured for the first time the single photon energy spectra of high energy
photons in the very forward region of proton-proton collisions at LHC.
After selecting data with common rapidity ranges, the two independent LHCf detectors
(Arm1 and Arm2) installed on either side of IP1 gave consistent results in two rapidity
ranges even though the geometrical acceptances of the two detectors differ.
The combined spectra of the two detectors are compared with the prediction of various
hadron interaction models.
It is found that none of the model predictions have perfect agreement with the
experimental results within statistical and systematic errors. 

The conservative systematic errors assigned in this analysis will be improved
upon in future studies.
In addition studies of other measurements like the inclusive single photon spectra,
inclusive single $\pi^{0}$ and neutron production spectra, and neutral particle
transverse momentum spectra are now ongoing using the already accumulated LHCf data.
By combining the LHCf data with the recent studies on particle production in the
central rapidity region of LHC collisions \cite{ref-LHCcentral}
it is now for the first time possible to make critical tests of hadron interaction
models by using collider data over a very wide rapidity range.

{\bf Acknowledgments}
We thank the CERN staff and the ATLAS collaboration for their essential contributions 
to the successful operation of LHCf.
Especially we appreciate continuous review of the experiment by Michelangelo Mangano,
Carsten Niebuhr and Mario Calvetti.
We also thank Tanguy Pierog for a quick verification of our MC calculations.
This work is partly supported by Grant-in-Aid for Scientific Research
by MEXT of Japan, the Mitsubishi Foundation in Japan and INFN in Italy.
The receipts of JSPS Research Fellowship (HM and TM), INFN fellowship for 
non Italian citizens (HM and KN) and the GCOE Program of Nagoya University 
"QFPU" from JSPS and MEXT of Japan (GM) are also acknowledged.
A part of this work was performed using the computer resource provided
by the Institute for the Cosmic-Ray Research (ICRR), University of Tokyo.

\clearpage

\end{document}